# Generation of nonlinear coherent states in a coherently pumped micromaser under intensity-dependent Jaynes-Cummings model

M. H. Naderi, M. Soltanolkotabi, and R. Roknizadeh
*Quantum Optics Group, Department of Physics, University of Isfahan, Isfahan, Iran*

**Abstract**

In this paper the possibility of generating nonlinear (*f*-deformed) coherent states of the radiation field in a micromaser is explored. It is shown that these states can be provided in a lossless micromaser cavity under the weak Jaynes-Cummings interaction with intensity–dependent coupling of large number of individually injected two-level atoms in a coherent superposition of the upper and lower states. In particular, we show that the so-called nonlinear squeezed vacuum and nonlinear squeezed first excited states, as well as the even and odd nonlinear coherent states can be generated in the presence of two-photon transitions.

## 1. Introduction

The importance of coherent states (CSs) of various Lie algebras in different branches of physics, particularly in quantum optics, hardly needs to be emphasized. Historically, the conventional CSs of the quantum harmonic oscillator corresponding to the Heisenberg-Weyl algebra were first introduced by Schroedinger [1].These states have quantum statistical properties like the classical radiation field and they define the limit between the classical and nonclassical behaviors, like squeezing, antibunching and sub-Poissonian statistics. Subsequently the notion was generalized in various ways. Motivations to generalize the concept have arisen from symmetry considerations [2,3], dynamics [4] and algebraic aspects [5].

Recently a generalized class of the conventional CSs called the nonlinear coherent states (NLCSs) [6] or the *f*-CSs [7], which can be classified as an algebraic generalization of the conventional CSs have been constructed. These states, which correspond to nonlinear algebras rather than Lie algebras, are defined as right eigenstates of the generalized annihilation operator $\hat{A} = \hat{a} f(\hat{N})$,

$$\hat{A}|z\rangle_f = z|z\rangle_f, \qquad (1)$$

where $f(\hat{N})$ is a reasonably well-behaved real function of the number operator $\hat{N} = \hat{a}^+ \hat{a}$ and $z$ is an arbitrary complex number. From (1) one can obtain an explicit form of the NLCSs in a number state representation

$$|z\rangle_f = C \sum_{n=0}^{\infty} z^n d_n |n\rangle, \qquad (2)$$

where the coefficients $d_n$'s and normalization constant *C*, respectively, are given by

$$d_0 = 1, \quad d_n = \left(\sqrt{n!} f(n)!\right)^{-1}, \quad f(n)! \equiv \prod_{j=1}^{n} f(j) \qquad (3)$$

$$C = \left(\sum_{n=0}^{\infty} d_n^2 |z|^{2n}\right)^{-1/2}. \qquad (4)$$

Actually, NLCSs have been known for many years under other names. The phase state[8] or its generalization [9] (known nowadays as the negative binomial state or the *SU*(1,1) group coherent states) and the photon-added CS [10] are two well-known examples of NLCSs. The physical meaning of NLCSs has been elucidated in [6,7], where it has been shown that such states may appear as stationary states of the center-of-mass motion of a trapped ion [6], or may be related to some nonlinear processes (such as a hypothetical "frequency blue shift " in high intensity photon beams [7] ). Furthermore, it has been shown that NLCSs exhibit various nonclassical features such as quadrature squeezing, number-phase squeezing and sub-Poissonian photon statistics[11].

On the other hand, in last few years the production and detection of nonclassical states of the radiation field have attracted a great deal of attention because of their latent applications in optical communication and in precision and sensitive measurements. One of the marvelous experimental and theoretical systems which can be utilized to produce nonclassical radiation is the one–atom maser or micromaser [12]. The system consists of a high-*Q* microwave cavity and a stream of Rydberg atoms which drive the field inside the cavity. The atomic beam is sufficiently sparse so that no more than one atom is in the cavity at any time. There are several schemes that have been proposed to produce number states [13], sub-Poissonian states [14], squeezed states [15] and trapping states [16] and the possibility of generating pure states of the field, the so-called tangent and cotangent states [17] has also been predicted by using micromasers. Although foregoing schemes have been discussed by utilizing the standard Jaynes-Cummings model [18], one may consider the intensity-dependent Jaynes-Cummings model. The intensity–dependent interaction is interesting since it represents a very simple case of a nonlinear interaction corresponding to a more realistic physical situation. Moreover, it can potentially provide various variants of the field state possessing interesting quantum statistical features.

In the present paper we study the possibility of generating various types of NLCSs in a resonant, lossless micromaser with injected two-level atoms in a coherent superposition of the upper and lower states. It is assumed that the atom-field coupling is intensity-dependent described by a generic function of the number operator, e.g. $f(\hat{N})$. By considering the weak interaction of large number of individual atoms and radiation through one as well as two-photon transitions we investigate the quantum evolution of the cavity-field state. In the next section we are dealing with the case of one-photon transitions and section 3 is devoted to the case of two-photon transitions. The results are summarized in section 4.

## 2. Quantum evolution of the cavity-field state: one-photon transitions

We consider a beam of monoenergetic two-level atoms injected into a lossless microwave cavity in which the atoms interact resonantly with the cavity mode for a finite time $\tau$. The micromaser is usually operated in the regime in which there is at most only one atom in the cavity at any time. It is assumed that the time of the interaction of each atom with the cavity-field is much shorter than the lifetime of all the atomic levels. Then the atomic spontaneous decay processes to other levels can be ignored while an atom is inside the cavity, which means that the joint evolution of the cavity-field and atoms is unitary.

We assume that injected atoms interact with the cavity-field through one-photon transitions and intensity-dependent coupling. The Hamiltonian describing the atom-field interaction in the rotating-wave approximation and in the interaction picture is given by the intensity-dependent Jaynes-Cummings model ($\hbar = 1$)

$$\hat{H}_I = g\left(\hat{A}|a\rangle\langle b| + \hat{A}^+|b\rangle\langle a|\right). \tag{5}$$

Here the atomic operator $|i\rangle\langle j|$ ($i \neq j$) denotes the transition from level $j$ to level $i$, $g$ is the coupling constant and $\hat{A}$ and $\hat{A}^+$ are constructed from the photon annihilation and creation operators $\hat{a}$ and $\hat{a}^+$ of the cavity–field,

$$\hat{A} = \hat{a}f(\hat{N}) \quad , \quad \hat{A}^+ = f(\hat{N})\hat{a}^+, \tag{6}$$

where the real function $f(\hat{N})$ describes the intensity dependence of atom-field interaction (it is assumed that the function $f(\hat{N})$ is such that it has no zeros at positive integer values of $n$, including zero). From the relations (6) it follows that $\hat{A}$, $\hat{A}^+$ and the number operator $\hat{N}$ satisfy the following closed algebraic relations

$$[\hat{A}, \hat{A}^+] = \{\hat{N}+1\}_f - \{\hat{N}\}_f \quad , \quad [\hat{N}, \hat{A}] = -\hat{A} \quad , \quad [\hat{N}, \hat{A}^+] = \hat{A}^+, \tag{7-a}$$

together with

$$\hat{A}|n\rangle = \sqrt{\{n\}_f}|n-1\rangle \quad , \quad \hat{A}^+|n\rangle = \sqrt{\{n+1\}_f}|n+1\rangle \quad , \quad \hat{N}|n\rangle = n|n\rangle, \tag{7-b}$$

where the symbol $\{X\}_f$ stands for $Xf^2(X)$. Thus the relations (7-a) represent a deformed Heisenberg algebra whose nature of deformation depends on the nonlinearity function $f(\hat{N})$. Clearly for $f(\hat{N})=1$ we regain the Heisenberg algebra.

It is easy to show that the corresponding time evolution operator $\hat{U}(\tau)$ can be expressed in the form

$$\hat{U}(\tau) \equiv \exp(-i\hat{H}_I\tau) = \cos(g\tau\sqrt{\{\hat{N}+1\}_f})|a\rangle\langle a| + \cos(g\tau\sqrt{\{\hat{N}\}_f})|b\rangle\langle b|$$
$$-i\frac{\sin(g\tau\sqrt{\{\hat{N}+1\}_f})}{\sqrt{\{\hat{N}+1\}_f}}\hat{A}|a\rangle\langle b| - i\frac{\sin(g\tau\sqrt{\{\hat{N}\}_f})}{\sqrt{\{\hat{N}\}_f}}\hat{A}^+|b\rangle\langle a| \qquad (8)$$

Let all atoms are initially prepared in a same superposition of the upper level $|a\rangle$ and the lower level $|b\rangle$. Therefore the initial density matrix of the $K^{th}$ atom can be written as

$$\hat{\rho}_A^{(K)}(t=0) = \sum_{i,j=a,b}\rho_{ij}|i_K\rangle\langle j_K|, \qquad (9)$$

where $\rho_{ii} \geq 0$, $\rho_{aa}+\rho_{bb}=1$, $\rho_{ab}=\rho_{ba}^*=|\rho_{ab}|\exp(i\varphi)$, $|\rho_{ab}|=|\rho_{ba}|\leq\sqrt{\rho_{aa}\rho_{bb}}$.

It should be noted that since there is a free evolution of the cavity-field density matrix in the time between the subsequent atoms entering the cavity, i.e., the matrix elements of the cavity-field density matrix acquire an extra phase factor $exp(i(n-n')\omega\delta t)$, where $\omega$ is the cavity resonance frequency and $\delta t$ is the time between the arrivals of subsequent atoms, we assume here that the time $\delta t$ is chosen in such a way that $\omega\delta t$ is equal to a multiple of $2\pi$. In this case the extra phase factor due to the free evolution is unity. Otherwise we should take it into account in the overall density matrix evolution. If the atoms were arriving at random times they would meet the cavity-field with random phases, and the cavity-field phase, which is associated with the non-diagonal elements of the density matrix, would necessarily become random (only diagonal elements would survive). This assumption is a very serious restriction to the model considered here. It means that atoms should be injected into the cavity in a well controllable way.

Under the above assumption the field density matrix, after passing $K$ atoms through the micromaser cavity reads as

$$\hat{\rho}_F^{(K)} = Tr_A\left(\hat{U}_K(\tau)\hat{\rho}_A^{(K)}\otimes\hat{\rho}_F^{(K-1)}\hat{U}^+(\tau)\right), \qquad (10)$$

in which $Tr_A$ indicates partial trace over the Hilbert space of the two-level atom. In writing (10) we have assumed that the state of the atom is not measured as it exits the cavity. The number of injected atoms $K$ is considered as a scaled evolution time of the system. By using (10) together with the expressions (8) and (9), we can easily get for the cavity-field density matrix elements the recursion relation

$\rho_F^{(K)}(n,n') = \langle n|\hat{\rho}_F^{(K)}|n'\rangle =$

$\left(\rho_{aa}\cos(g\tau\sqrt{\{n+1\}_f})\cos(g\tau\sqrt{\{n'+1\}_f}) + \rho_{bb}\cos(g\tau\sqrt{\{n\}_f})\cos(g\tau\sqrt{\{n'\}_f})\right)\rho_F^{(K-1)}(n,n')$

$\quad + \rho_{bb}\sin(g\tau\sqrt{\{n+1\}_f})\sin(g\tau\sqrt{\{n'+1\}_f})\rho_F^{(K-1)}(n+1,n'+1)$

$\quad + \rho_{aa}\sin(g\tau\sqrt{\{n\}_f})\sin(g\tau\sqrt{\{n'\}_f})\rho_F^{(K-1)}(n-1,n'-1)$

$$+i|\rho_{ab}|exp(i\varphi)\cos(g\tau\sqrt{\{n+1\}_f})\sin(g\tau\sqrt{\{n'+1\}_f})\rho_F^{(K-1)}(n,n'+1)$$
$$+i|\rho_{ab}|exp(-i\varphi)\cos(g\tau\sqrt{\{n\}_f})\sin(g\tau\sqrt{\{n'\}_f})\rho_F^{(K-1)}(n,n'-1)$$
$$-i|\rho_{ab}|exp(-i\varphi)\sin(g\tau\sqrt{\{n+1\}_f})\cos(g\tau\sqrt{\{n'+1\}_f})\rho_F^{(K-1)}(n+1,n')$$
$$-i|\rho_{ab}|exp(i\varphi)\sin(g\tau\sqrt{\{n\}_f})\cos(g\tau\sqrt{\{n'\}_f})\rho_F^{(K-1)}(n-1,n'),$$
(11)

with the initial condition $\rho_F^{(K=0)}(n,n') = \rho_F^{(0)}(n,n')$. It is seen from the recursion relation (11) that the coupling between the diagonal matrix elements $\rho_F^{(K)}(n,n)$ and the off-diagonal elements $\rho_F^{(K)}(n,n\pm 1) = \rho_F^{(K)*}(n\pm 1,n)$ occurs only when the atomic coherence $\rho_{ab}$ is present. If the micromaser is pumped by unpolarized atoms ($\rho_{ab}=0$) then the off-diagonal elements don't occur, and consequently the field phase is always random. However, atoms prepared in a coherent superposition of their states before entering the micromaser cavity create nonvanishing off-diagonal elements, that is they create a preferred phase field [19].

In order to solve the recursion relation (11) we adopt a method which has firstly been used by Kien *et al* [20]. We introduce the phase-independent matrix elements $\widetilde{\rho}^{(K)}(n,n')$ through the definition

$$\rho_F^{(K)}(n,n') = (i\exp(-i\varphi))^{n'-n}\left(\prod_{\ell=1}^{n}\sin(g\tau\sqrt{\{\ell\}_f})\prod_{\ell'=1}^{n'}\sin(g\tau\sqrt{\{\ell'\}_f})\right)\rho_{aa}^{(n+n')/2}\widetilde{\rho}^{(K)}(n,n').$$
(12)

Substituting this expression into (11), we get the recursion relation
$$\widetilde{\rho}^{(K)}(n,n') =$$
$$\left(\rho_{aa}\cos(g\tau\sqrt{\{n+1\}_f})\cos(g\tau\sqrt{\{n'+1\}_f}) + \rho_{bb}\cos(g\tau\sqrt{\{n\}_f})\cos(g\tau\sqrt{\{n'\}_f})\right)\widetilde{\rho}^{(K-1)}(n,n')$$
$$+\rho_{aa}\rho_{bb}\sin^2(g\tau\sqrt{\{n+1\}_f})\sin^2(g\tau\sqrt{\{n'+1\}_f})\widetilde{\rho}^{(K-1)}(n+1,n'+1)$$
$$+\widetilde{\rho}^{(K-1)}(n-1,n'-1)$$
$$-\rho_{aa}\sqrt{\rho_{bb}}\cos(g\tau\sqrt{\{n+1\}_f})\sin^2(g\tau\sqrt{\{n'+1\}_f})\widetilde{\rho}^{(K-1)}(n,n'+1)$$
$$-\rho_{aa}\sqrt{\rho_{bb}}\sin^2(g\tau\sqrt{\{n+1\}_f})\cos(g\tau\sqrt{\{n'+1\}_f})\widetilde{\rho}^{(K-1)}(n+1,n')$$
$$+\sqrt{\rho_{bb}}\cos(g\tau\sqrt{\{n\}_f})\widetilde{\rho}^{(K-1)}(n,n'-1) + \sqrt{\rho_{bb}}\cos(g\tau\sqrt{\{n'\}_f})\widetilde{\rho}^{(K-1)}(n-1,n'),$$
(13)

with the initial condition
$$\widetilde{\rho}^{(0)}(n,n') = (i\exp(-i\varphi))^{n-n'}\left(\prod_{\ell=1}^{n}\sin(g\tau\sqrt{\{\ell\}_f})\prod_{\ell'=1}^{n'}\sin(g\tau\sqrt{\{\ell'\}_f})\right)^{-1}\rho_{aa}^{-(n+n')/2}\rho_F^{(0)}(n,n').$$
(14)

Now we consider the case of weak atom-field interaction, that is,
$$g\tau \ll 1 \quad , \quad g\tau\sqrt{\{\overline{n}\}} \equiv g\tau\sqrt{\overline{n}f^2(\overline{n})} \ll 1,$$
(15)

where $\overline{n}$ is the mean photon number of the cavity-field. In addition we assume that the variance of the photon-number distribution always to be not too large for all the time. In the first-order approximation that is,
$$\sin(g\tau\sqrt{\{n\}}) \approx g\tau\sqrt{\{n\}} \quad , \sin^2(g\tau\sqrt{\{n\}}) \approx 0 \quad , \cos(g\tau\sqrt{\{n\}}) \approx 1$$

the recursion relation (13) becomes

$$\tilde{\rho}^{(K)}(n,n') \approx \tilde{\rho}^{(K-1)}(n,n') + \tilde{\rho}^{(K-1)}(n-1,n'-1) + \sqrt{\rho_{bb}}\left(\tilde{\rho}^{(K-1)}(n,n'-1) + \tilde{\rho}^{(K-1)}(n-1,n')\right)$$
(16)

The solution of Eq.(16) is easily found to be

$$\tilde{\rho}^{(K)}(n,n') \approx \sum_{k,k'=0}^{K} \sum_{p=0}^{\min(k,k')} \frac{K!\rho_{bb}^{(k+k'-2p)/2}}{p!(k-p)!(k'-p)!(K-k-k'+p)!} \tilde{\rho}^{(0)}(n-k,n'-k').$$
(17)

Since the expression (17) is not very convenient to be used for large $K$, we prefer to use the truncated form

$$\tilde{\rho}^{(K)}(n,n') \approx \sum_{k=0}^{n} \sum_{k'=0}^{n'} \sum_{p=0}^{\min(k,k')} \frac{K!\rho_{bb}^{(k+k'-2p)/2}}{p!(k-p)!(k'-p)!(K-k-k'+p)!} \tilde{\rho}^{(0)}(n-k,n'-k').$$
(18)

Substituting Eq.(18) into Eq.(12), we obtain for $\rho_F^{(K)}(n,n')$ the approximate expression

$$\rho_F^{(K)}(n,n') \approx (i\exp(-i\varphi))^{n'-n}(g\tau\sqrt{\rho_{aa}})^{n+n'}$$

$$\times \sqrt{\{n\}_f!\{n'\}_f!}\sum_{k=0}^{n}\sum_{k'=0}^{n'}\sum_{p=0}^{\min(k,k')} \frac{K!\rho_{bb}^{(k+k'-2p)/2}}{p!(k-p)!(k'-p)!(K-k-k'+p)!} \tilde{\rho}^{(0)}(n-k,n'-k'),$$
(19)

where by definition $\{n\}_f! = \{n\}_f\{n-1\}_f\{n-2\}_f....1$ and $\{0\}_f!=1$. Now let $\rho_{bb} \neq 0$ and $K \gg 1$. The relation between $(p+1)^{\text{th}}$ and $p^{\text{th}}$ terms in the sum on the r.h.s of expression (18) is

$$\frac{(k-p)(k'-p)}{\rho_{bb}(p+1)(K-k-k'+1)} \leq \frac{kk'}{\rho_{bb}(K-k-k')}.$$
(20)

As it is seen, in the region of values of $n$ and $n'$ such that $n+n'+\frac{nn'}{\rho_{bb}} \ll K$ the term with $p=0$ in the sum on the r.h.s of (18) dominates. Keeping only the $p=0$ term and using the approximation $K!/(K-k-k')! \approx K^{(k+k')}$, from expression (19) we find

$$\rho_F^{(K)}(n,n') \approx$$

$$\sqrt{\{n\}_f!\{n'\}_f!}\sum_{k=0}^{n}\sum_{k'=0}^{n'}(i\exp(-i\varphi))^{k'-k}\frac{(Kg\tau\sqrt{\rho_{aa}\rho_{bb}})^{k+k'}}{k!k'!\sqrt{\{n-k\}_f!\{n'-k'\}_f!}}\rho_F^{(0)}(n-k,n'-k'),$$
(21)

in which we have made use of (14). The expression (21) gives the matrix elements of the cavity-field density matrix after passing a large number of injected atoms through the micromaser cavity where each atom undergoes one-photon transitions under the weak atom-field interaction with intensity-dependent coupling. If the micromaser starts from a pure state $\left|\psi_F^{(0)}\right\rangle$, then the cavity-field evolves into the pure state $\left|\psi_F^{(K)}\right\rangle$ as follows

$$\left\langle n\middle|\psi_F^{(K)}\right\rangle \approx \sum_{k=0}^{n} \frac{z^k}{k!\sqrt{\{n-k\}_f!}}\sqrt{\{n\}_f!}\left\langle n-k\middle|\psi_F^{(0)}\right\rangle$$
(22)

with $z = -i\exp(-i\varphi)Kg\tau\sqrt{\rho_{aa}\rho_{bb}}$.

Now let us assume that the cavity-field is initially in the vacuum state, i.e., $\left|\psi_F^{(0)}\right\rangle = \left|0\right\rangle$. Thus from (22) we find the following approximate expression for the cavity-field state after passing $K$ atoms ($K \gg 1$)

$$\left|\Psi_F^{(K)}\right\rangle \equiv |z\rangle_f{}' = C'\sum_{n=0}^{\infty}\frac{z^n}{n!}\sqrt{\{n\}_f!}|n\rangle = C'\sum_{n=0}^{\infty}z^n\, d'_n\,|n\rangle, \qquad (23)$$

where

$$d'_0 = 1, \quad d'_n = \left(\sqrt{n!}/f(n)!\right)^{-1}, \qquad (24\text{-a})$$

and $C'$ is the normalization constant given by

$$C' = \left(\sum_{n=0}^{\infty} d'^2_n\, |z|^{2n}\right)^{-1/2}. \qquad (24\text{-b})$$

It is evident that for $f(n) = 1$ the state vector (23) describes the usual CS. Therefore it is reasonable to interpret the state (23) as a NLCS of the cavity-field. But, since the expansion coefficients $d'_n$ are different from the coefficients $d_n$, given by (3) (provided of course we use the same nonlinearity function $f(n)$ in both the cases) the NLCS obtained in (23) is distinct from the NLCS defined in Eq.(2).

In order to get more clear insight to the above result we present the following argument. From the relation (7-a) we find that the r.h.s of the commutator $[\hat{A},\hat{A}^+]$ is a nonlinear function of the number operator $\hat{N}$. As a result BCH disentangling theorem [21] can not be applied and one can not use the displacement operator $\exp(z\hat{A}-z^*\hat{A}^+)$ to construct coherent states. Therefore one may seek for an operator $\hat{B}^+$ which is conjugate of the operator $\hat{A}$, that is $[\hat{A},\hat{B}^+]=1$ while their Hermitian conjugates $\hat{A}^+$ and $\hat{B}$ satisfy the dual algebra $[\hat{B},\hat{A}^+]=1$. From (6) it is easily found that

$$\hat{B} = \hat{a}\frac{1}{f(\hat{N})}, \qquad \hat{B}^+ = \frac{1}{f(\hat{N})}\hat{a}^+. \qquad (25)$$

Let us now consider the following displacement operators

$$\hat{D}_f(z) = \exp(z\hat{B}^+ - z^*\hat{A}), \qquad \hat{D}'_f(z) = \exp(z\hat{A}^+ - z^*\hat{B}) \qquad (26)$$

and note that for any two operators $\hat{X}$ and $\hat{Y}$ satisfying the relation $[\hat{X},\hat{Y}]=1$ the BCH theorem results in

$$\exp(z\hat{X} - z^*\hat{Y}) = \exp(-zz^*/2)\exp(z\hat{X})\exp(-z^*\hat{Y}). \qquad (27)$$

Now it is easy to find that the NLCSs (2) can be defined as $|z\rangle_f = \hat{D}_f(z)|0\rangle$, corresponding to the dual algebra $[\hat{B},\hat{A}^+]=1$ while the NLCSs given by (23) can be defined as $|z\rangle_f{}' = \hat{D}'_f(z)|0\rangle$, corresponding to the algebra $[\hat{A},\hat{B}^+]=1$. In addition, we have

$$\hat{A}|z\rangle_f = \hat{a}f(\hat{N})|z\rangle_f = z|z\rangle_f, \qquad \hat{B}|z\rangle_f{}' = \hat{a}\frac{1}{f(\hat{N})}|z\rangle_f{}' = z|z\rangle_f{}'. \qquad (28)$$

In this manner, we conclude that the intensity-dependent Jaynes-Cummings Hamiltonian (5), under the conditions of weak atom-field interaction and the passage of large number of polarized atoms through the cavity initially prepared in vacuum state, results in the NLCS (23) which is the eigenstates of the deformed annihilation operator $\hat{B}$. While if one considers the intensity-dependent interaction Hamiltonian as $\hat{H}_I = g\left(\hat{B}|a\rangle\langle b| + \hat{B}^+|b\rangle\langle a|\right)$ then under the same conditions the cavity-field evolves to the NLCS (2) which is the eigenstate of the operator $\hat{A}$.

## 3. Quantum evolution of the cavity-field state: two-photon transitions

So far, we have discussed the possibility of the generation NLCSs in a micromaser in which the injected atoms interact with the cavity–field through one-photon transitions. Now we examine the problem for the case of two-photon transitions. For this purpose let us consider the following Jaynes-Cummings Hamiltonian

$$\hat{H}_I = g\left(\hat{C}|a\rangle\langle b| + \hat{C}^+|b\rangle\langle a|\right), \tag{29}$$

where

$$\hat{C} = \hat{a}^2 f(\hat{N}) \quad , \quad \hat{C}^+ = f(\hat{N})(\hat{a}^+)^2. \tag{30}$$

In fact the Hamiltonian (29) describes the two-photon intensity-dependent atom-field coupling. The operators $\hat{C}, \hat{C}^+$ and $\hat{N}$ satisfy the following closed algebraic relations

$$[\hat{C},\hat{C}^+] = [\hat{N}+2]_f - [\hat{N}]_f \quad , \quad [\hat{N},\hat{C}] = -2\hat{C} \quad , \quad [\hat{N},\hat{C}^+] = 2\hat{C}^+, \tag{31-a}$$

together with

$$\hat{A}|n\rangle = \sqrt{[n]_f}|n-2\rangle \quad , \quad \hat{A}^+|n\rangle = \sqrt{[n+2]_f}|n+2\rangle \quad , \quad \hat{N}|n\rangle = n|n\rangle, \tag{31-b}$$

in which the symbol $[X]_f$ stands for $X(X-1)f^2(X)$. It should be noted that the deformed operator $\hat{C}$ annihilates both the vacuum state $|0\rangle$ and first excited state $|1\rangle$.

In this case we obtain the following recursion relation for the cavity-field density matrix elements after passing $K$ polarized atoms through the micromaser cavity

$$\rho_F^{(K)}(n,n') = \langle n|\hat{\rho}_F^{(K)}|n'\rangle =$$

$$\left(\rho_{aa}\cos(g\tau\sqrt{[n+2]_f})\cos(g\tau\sqrt{[n'+2]_f}) + \rho_{bb}\cos(g\tau\sqrt{[n]_f})\cos(g\tau\sqrt{[n']_f})\right)\rho_F^{(K-1)}(n,n')$$

$$+ \rho_{bb}\sin(g\tau\sqrt{[n+2]_f})\sin(g\tau\sqrt{[n'+2]_f})\rho_F^{(K-1)}(n+2,n'+2)$$

$$+ \rho_{aa}\sin(g\tau\sqrt{[n]_f})\sin(g\tau\sqrt{[n']_f})\rho_F^{(K-1)}(n-2,n'-2)$$

$$+ i|\rho_{ab}|\exp(i\varphi)\cos(g\tau\sqrt{[n+2]_f})\sin(g\tau\sqrt{[n'+2]_f})\rho_F^{(K-1)}(n,n'+2)$$

$$+ i|\rho_{ab}|\exp(-i\varphi)\cos(g\tau\sqrt{[n]_f})\sin(g\tau\sqrt{[n']_f})\rho_F^{(K-1)}(n,n'-2)$$

$$- i|\rho_{ab}|\exp(-i\varphi)\sin(g\tau\sqrt{[n+2]_f})\cos(g\tau\sqrt{[n'+2]_f})\rho_F^{(K-1)}(n+2,n')$$

$$- i|\rho_{ab}|\exp(i\varphi)\sin(g\tau\sqrt{[n]_f})\cos(g\tau\sqrt{[n']_f})\rho_F^{(K-1)}(n-2,n'). \tag{32}$$

To solve the above equation we employ the same method as in pervious section. A moment's inspection of the equation (32) shows that it is convenient to propose the phase-independent matrix elements $\tilde{\rho}^{(K)}(n,n')$ through the definitions

$$\rho_F^{(K)}(n,n') = (i\exp(-i\varphi))^{(n'-n)/2}\left(\prod_{\ell=1}^{n/2}\sin(g\tau\sqrt{[2\ell]_f})\prod_{\ell'=1}^{n'/2}\sin(g\tau\sqrt{[2\ell']_f})\right)\rho_{aa}^{(n+n')/4}\tilde{\rho}^{(K)}(n,n') \tag{33-a}$$

for even $n,n'$,

$$\rho_F^{(K)}(n,n') =$$

$$(i\exp(-i\varphi))^{(n'-n)/2}\left(\prod_{\ell=1}^{(n-1)/2}\sin(g\tau\sqrt{[2\ell+1]_f})\prod_{\ell'=1}^{(n'-1)/2}\sin(g\tau\sqrt{[2\ell'+1]_f})\right)\rho_{aa}^{(n+n')/4}\tilde{\rho}^{(K)}(n,n') \tag{33-b}$$

for odd $n,n'$,

$$\rho_F^{(K)}(n,n') =$$
$$(i\exp(-i\varphi))^{(n'-n)/2}\left(\prod_{\ell=1}^{n/2}\sin(g\tau\sqrt{[2\ell]_f})\prod_{\ell'=1}^{(n'-1)/2}\sin(g\tau\sqrt{[2\ell'+1]_f})\right)\rho_{aa}^{(n+n')/4}\tilde{\rho}^{(K)}(n,n'),$$

(33-c)

for even $n$ and odd $n'$ and
$$\rho_F^{(K)}(n,n') =$$
$$(i\exp(-i\varphi))^{(n'-n)/2}\left(\prod_{\ell=1}^{(n-1)/2}\sin(g\tau\sqrt{[2\ell+1]_f})\prod_{\ell'=1}^{n'/2}\sin(g\tau\sqrt{[2\ell']_f})\right)\rho_{aa}^{(n+n')/4}\tilde{\rho}^{(K)}(n,n'),$$

(33-d)

for odd $n$ and even $n'$. By substituting the expressions (33) in (32) we find the following recursion relation for the matrix elements $\tilde{\rho}^{(K)}(n,n')$ which is valid for all values of $n$ and $n'$

$$\tilde{\rho}^{(K)}(n,n') =$$
$$\left(\rho_{aa}\cos(g\tau\sqrt{[n+2]_f})\cos(g\tau\sqrt{[n'+2]_f})+\rho_{bb}\cos(g\tau\sqrt{[n]_f})\cos(g\tau\sqrt{[n']_f})\right)\tilde{\rho}^{(K-1)}(n,n')$$
$$+\rho_{aa}\rho_{bb}\sin^2(g\tau\sqrt{[n+2]_f})\sin^2(g\tau\sqrt{[n'+2]_f})\tilde{\rho}^{(K-1)}(n+2,n'+2)$$
$$+\tilde{\rho}^{(K-1)}(n-2,n'-2)$$
$$-\rho_{aa}\sqrt{\rho_{bb}}\cos(g\tau\sqrt{[n+2]_f})\sin^2(g\tau\sqrt{[n'+2]_f})\tilde{\rho}^{(K-1)}(n,n'+2)$$
$$-\rho_{aa}\sqrt{\rho_{bb}}\sin^2(g\tau\sqrt{[n+2]_f})\cos(g\tau\sqrt{[n'+2]_f})\tilde{\rho}^{(K-1)}(n+2,n')$$
$$+\sqrt{\rho_{bb}}\cos(g\tau\sqrt{[n]_f})\tilde{\rho}^{(K-1)}(n,n'-2)+\sqrt{\rho_{bb}}\cos(g\tau\sqrt{[n']_f})\tilde{\rho}^{(K-1)}(n-2,n'),$$

(34)

with the initial conditions

$$\tilde{\rho}^{(0)}(n,n') = (i\exp(-i\varphi))^{(n-n')/2}\left(\prod_{\ell=1}^{n/2}\sin(g\tau\sqrt{[2\ell]_f})\prod_{\ell'=1}^{n'/2}\sin(g\tau\sqrt{[2\ell']_f})\right)^{-1}\rho_{aa}^{-(n+n')/4}\rho_F^{(0)}(n,n')$$

(35-a)

for even $n,n'$,
$$\tilde{\rho}^{(0)}(n,n') =$$
$$(i\exp(-i\varphi))^{(n-n')/2}\left(\prod_{\ell=1}^{(n-1)/2}\sin(g\tau\sqrt{[2\ell+1]_f})\prod_{\ell'=1}^{(n'-1)/2}\sin(g\tau\sqrt{[2\ell'+1]_f})\right)^{-1}\rho_{aa}^{-(n+n')/4}\rho_F^{(0)}(n,n')$$

(35-b)

for odd $n,n'$,
$$\tilde{\rho}^{(0)}(n,n') =$$
$$(i\exp(-i\varphi))^{(n-n')/2}\left(\prod_{\ell=1}^{n/2}\sin(g\tau\sqrt{[2\ell]_f})\prod_{\ell'=1}^{(n'-1)/2}\sin(g\tau\sqrt{[2\ell'+1]_f})\right)^{-1}\rho_{aa}^{-(n+n')/4}\rho_F^{(0)}(n,n')$$

(35-c)

for even $n$ and odd $n'$ and
$$\tilde{\rho}^{(0)}(n,n') =$$
$$(i\exp(-i\varphi))^{(n-n')/2}\left(\prod_{\ell=1}^{(n-1)/2}\sin(g\tau\sqrt{[2\ell+1]_f})\prod_{\ell'=1}^{n'/2}\sin(g\tau\sqrt{[2\ell']_f})\right)^{-1}\rho_{aa}^{-(n+n')/4}\rho_F^{(0)}(n,n').$$

(35-d)

for odd $n$ and even $n'$.

Now, as before, we consider the case of weak atom-field interaction, that is,

$$g\tau \ll 1 \ , \ g\tau\sqrt{[n]} \equiv g\tau\sqrt{n(n-1)f^2(n)} \ll 1. \tag{36}$$

In the first-order approximation the recursion relation (34) becomes

$$\tilde{\rho}^{(K)}(n,n') \approx \tilde{\rho}^{(K-1)}(n,n') + \tilde{\rho}^{(K-1)}(n-2,n'-2) + \sqrt{\rho_{bb}}\left(\tilde{\rho}^{(K-1)}(n,n'-2) + \tilde{\rho}^{(K-1)}(n-2,n')\right) \tag{37}$$

whose solution reads as

$$\tilde{\rho}^{(K)}(n,n') \approx \sum_{k,k'=0}^{K} \sum_{p=0}^{\min(k,k')} \frac{K! \rho_{bb}^{(k+k'-2p)/2}}{p!(k-p)!(k'-p)!(K-k-k'+p)!} \tilde{\rho}^{(0)}(n-2k, n'-2k'). \tag{38}$$

Following the same lines as those of the previous section we find that the micromaser field, initially prepared in a pure state $|\psi_F^{(0)}\rangle$, evolves into the pure state $|\psi_F^{(K)}\rangle$ as follows

$$\langle n|\psi_F^{(K)}\rangle = \sum_{k=0}^{n} \frac{z^k}{k!\sqrt{[n-2k]_f!!}}\sqrt{[n]_f!!}\langle n-2k|\psi_F^{(0)}\rangle, \tag{39}$$

where $z = -i\exp(-i\varphi)Kg\tau\sqrt{\rho_{aa}\rho_{bb}}$ and by definition

$$[n]!! = [2][4][6]....[n-2][n], \text{ for } n=2m \ (m=0,1,2,...),$$
$$[n]!! = [3][5][7]....[n-2][n], \text{ for } n=2m+1 \ (m=0,1,2,...) \tag{40}$$

and $[0]_f!! = [1]_f!! = 1$.

Now let us assume that the cavity-field is initially in the vacuum state, i.e., $|\psi_F^{(0)}\rangle = |0\rangle$. Thus from (39) we find the following approximate expression for the cavity-field state after passing $K$ atoms ($K \gg 1$)

$$|\psi_F^{(K)}\rangle \equiv |z,0\rangle_f = C_0 \sum_{m=0}^{\infty} \frac{z^m}{m!}\sqrt{(2m)!}\ f(2m)!!|2m\rangle, \tag{41-a}$$

where $f(2m)!! = f(2)f(4)...f(2m-2)f(2m)$ and $C_0$ is the normalization constant given by

$$C_0 = \left(\sum_{m=0}^{\infty}|z|^{2m}(2m)!(f(2m)!!)^2/(m!)^2\right)^{-1/2}. \tag{41-b}$$

If the cavity-field starts from the first excited state, $|\psi_F^{(0)}\rangle = |1\rangle$, then it evolves to the state

$$|\psi_F^{(K)}\rangle \equiv |z,1\rangle_f = C_1 \sum_{m=0}^{\infty} \frac{z^m}{m!}\sqrt{(2m+1)!}\ f(2m+1)!!|2m+1\rangle, \tag{42-a}$$

where $f(2m+1)!! = f(3)f(5)...f(2m-1)f(2m+1)$ and $C_1$ is the normalization constant given by

$$C_1 = \left(\sum_{m=0}^{\infty}|z|^{2m}(2m+1)!(f(2m+1)!!)^2/(m!)^2\right)^{-1/2}. \tag{42-b}$$

In order to make more clear the nature of the states (41) and (42), we present the following argument. As stated before there are two vacua for the operator $\hat{C}$, namely $|0\rangle$ and $|1\rangle$. By applying the method of Shantha *et al* [22] one can construct the operator $\hat{B}_i^+$ ($i = 0,1$) such that $[\hat{C}, \hat{B}_i^+]=1$. For the sector $S_0$, constructed by repeatedly applying $\hat{C}^+$ on the ground state $|0\rangle$, we obtain

$$\hat{B}_0^+ = \frac{1}{2}(\hat{a}^+)^2 \frac{1}{\hat{N}+1} \frac{1}{f(\hat{N}+2)} \quad ; \quad \left([\hat{C}, \hat{B}_0^+]=1, [\hat{B}_0, \hat{C}^+]=1\right) \tag{43}$$

and for the sector $S_1$, constructed by repeatedly applying $\hat{C}^+$ on the first excited state $|1\rangle$, we obtain

$$\hat{B}_1^+ = \frac{1}{2}(\hat{a}^+)^2 \frac{1}{\hat{N}+2} \frac{1}{f(\hat{N}+2)} \quad ; \quad \left([\hat{C}, \hat{B}_1^+]=1, [\hat{B}_1, \hat{C}^+]=1\right). \tag{44}$$

We now construct the two following displacement operators corresponding to the algebras $[\hat{B}_0, \hat{C}^+]=1$ and $[\hat{B}_1, \hat{C}^+]=1$, respectively

$$\hat{D}_f^{(0)}(z) = \exp(z\hat{C}^+ - z^*\hat{B}_0) = \exp(-zz^*/2)\exp(z\hat{C}^+)\exp(-z^*\hat{B}_0), \tag{45-a}$$

$$\hat{D}_f^{(1)}(z) = \exp(z\hat{C}^+ - z^*\hat{B}_1) = \exp(-zz^*/2)\exp(z\hat{C}^+)\exp(-z^*\hat{B}_1). \tag{45-b}$$

Applying $\hat{D}_f^{(0)}(z)$ on $|0\rangle$ we obtain

$$\exp(z\hat{C}^+ - z^*\hat{B}_0)|0\rangle = \sum_{m=0}^{\infty} \frac{z^m}{m!}(f(\hat{N})(\hat{a}^+)^2)^m |0\rangle$$

$$= \sum_{m=0}^{\infty} \frac{z^m}{m!}\sqrt{(2m)!} f(2m)!! |2m\rangle \tag{46}$$

which is the same as the state $|z,0\rangle_f$ up to a normalization constant. On the other hand applying the operator $\hat{D}_f^{(1)}(z)$ on the state $|1\rangle$ yields

$$\exp(z\hat{C}^+ - z^*\hat{B}_1)|1\rangle = \sum_{m=0}^{\infty} \frac{z^m}{m!}(f(\hat{N})(\hat{a}^+)^2)^m |1\rangle$$

$$= \sum_{m=0}^{\infty} \frac{z^m}{m!}\sqrt{(2m+1)!} f(2m+1)!! |2m+1\rangle \tag{47}$$

which is the same as the state $|z,1\rangle_f$ up to a normalization constant. Additionally, it is easy to examine that

$$\hat{B}_0|z,0\rangle_f = z|z,0\rangle_f \quad , \quad \hat{B}_1|z,1\rangle_f = z|z,1\rangle_f \tag{48}$$

So each of the two states $|z,0\rangle_f$ and $|z,1\rangle_f$ not only can be obtained by the application of a displacement type operator but also as a nonlinear ( or $f$-deformed) annihilation operator eigenstate. In this manner each of these states can be interpreted as a type of nonlinear coherent states. Besides, the structures of the states $|z,0\rangle_f$ and $|z,1\rangle_f$ are remindful of the usual squeezed vacuum and squeezed first excited states, respectively [23] in the limit $f(n)=1$. Accordingly, it is reasonable to consider the states $|z,0\rangle_f$ and $|z,1\rangle_f$, respectively, as the nonlinear ($f$-deformed) squeezed vacuum and nonlinear ($f$-deformed) squeezed first excited states. These two states correspond to the algebras $[\hat{B}_0, \hat{C}^+]=1$ and $[\hat{B}_1, \hat{C}^+]=1$, respectively.

Therefore, if the atom-field interaction is described by the two-photon intensity-dependent Jaynes-Cummings Hamiltonian (29), then under the conditions of no losses and weak atom-field interaction together with a large enough of polarized injected atoms two other types of NLCSs can be created.

Finally, let us consider the two cases in which the atom-field interaction is governed by the operators $(\hat{B}_0, \hat{B}_0^+)$ and $(\hat{B}_1, \hat{B}_1^+)$, respectively.

In the first case the corresponding interaction Hamiltonian is

$$\hat{H}_I = g\left(\hat{B}_0 |a\rangle\langle b| + \hat{B}_0^+ |b\rangle\langle a|\right). \tag{49}$$

The operators $\hat{B}_0, \hat{B}_0^+$ and $\hat{N}$ satisfy the following closed algebraic relations

$$[\hat{B}_0, \hat{B}_0^+] = [[\hat{N}+2]]_f^{(0)} - [[\hat{N}]]_f^{(0)} \quad , \quad [\hat{N}, \hat{B}_0] = -2\hat{B}_0 \quad , \quad [\hat{N}, \hat{B}_0^+] = 2\hat{B}_0^+, \tag{50-a}$$

together with

$$\hat{B}_0 |n\rangle = \sqrt{[[n]]_f^{(0)}} |n-2\rangle \quad , \quad \hat{B}_0^+ |n\rangle = \sqrt{[[n+2]]_f^{(0)}} |n+2\rangle \quad , \quad \hat{N}|n\rangle = n|n\rangle, \tag{50-b}$$

where the symbol $[[X]]_f^{(0)}$ stands for $\dfrac{1}{4}\dfrac{X}{X-1}\dfrac{1}{f^2(X)}$. By applying exactly the same procedure as before we find that the micromaser field, initially in the vacuum state $|0\rangle$, after passing a large number of injected atoms evolves to the state

$$|\psi_F^{(K)}\rangle \equiv |z,0\rangle_f^{(e)} = C_e \sum_{m=0}^{\infty} \frac{z^m}{\sqrt{(2m)!}\, f(2m)!!} |2m\rangle \tag{51-a}$$

with

$$C_e = \left(\sum_{m=0}^{\infty} |z|^{2m} / (2m)!(f(2m)!!)^2 \right)^{-1/2} \tag{51-b}$$

as the normalization constant. Furthermore, it is easy to show that the state $|z,0\rangle_f^{(e)}$ can be constructed either by the action of the displacement operator $\hat{D}_f^{(0)'}(\hat{z}) = \exp(z\hat{B}_0^+ - z^*\hat{C})$ on the ground state $|0\rangle$ or as the eigenstate of the operator $\hat{C}$ with the eigenvalue $z$.

Let us now turn to the second case. The corresponding interaction Hamiltonian reads as

$$\hat{H}_I = g\left(\hat{B}_1 |a\rangle\langle b| + \hat{B}_1^+ |b\rangle\langle a|\right). \tag{52}$$

The operators $\hat{B}_1, \hat{B}_1^+$ and $\hat{N}$ satisfy the following closed algebraic relations

$$[\hat{B}_1, \hat{B}_1^+] = [[\hat{N}+2]]_f^{(1)} - [[\hat{N}]]_f^{(1)} \quad , \quad [\hat{N}, \hat{B}_1] = -2\hat{B}_1 \quad , \quad [\hat{N}, \hat{B}_1^+] = 2\hat{B}_1^+, \tag{53-a}$$

together with

$$\hat{B}_1 |n\rangle = \sqrt{[[n]]_f^{(1)}} |n-2\rangle \quad , \quad \hat{B}_1^+ |n\rangle = \sqrt{[[n+2]]_f^{(1)}} |n+2\rangle \quad , \quad \hat{N}|n\rangle = n|n\rangle, \tag{53-b}$$

where the symbol $[[X]]_f^{(1)}$ stands for $\dfrac{1}{4}\dfrac{X-1}{X}\dfrac{1}{f^2(X)}$. This time we find that the micromaser field, initially prepared in the first excited state $|1\rangle$, after passing a large number of injected atoms evolves to the state

$$|\psi_F^{(K)}\rangle \equiv |z,1\rangle_f^{(o)} = C_o \sum_{m=0}^{\infty} \frac{z^m}{\sqrt{(2m+1)!}\, f(2m+1)!!} |2m+1\rangle \tag{54-a}$$

with

$$C_o = \left( \sum_{m=0}^{\infty} |z|^{2m} / (2m+1)!(f(2m+1)!!)^2 \right)^{-1/2}. \tag{54-b}$$

as the normalization constant. In addition, it is found that the state $|z,1\rangle_f^{(o)}$ can be constructed either by the action of the displacement operator $\hat{D}_f^{(1)'}(\hat{z}) = \exp(z\hat{B}_1^+ - z^*\hat{C})$ on the state $|1\rangle$ or as the eigenstate of the operator $\hat{C}$ with the eigenvalue $z$.

In fact, the states $|z,0\rangle_f^{(e)}$ and $|z,1\rangle_f^{(o)}$ are respectively the even and odd NLCSs [24] which are defined as the extensions of the notions of usual even and odd coherent states. Here we point out that, as another production scheme, it has been suggested that a class of these states with a specific form of $f(\hat{N})$ can be generated in the center-of-mass motion of a trapped and bichromatically laser-driven two-level ion [25].

Therefore we have shown the possibility of the generation of even and odd NLCSs of the radiation field in a coherently pumped micromaser where the atom-field interaction is governed by (49) and (52), respectively. Algebraically the two states correspond to the dual algebras $[\hat{C}, \hat{B}_0^+] = 1$ and $[\hat{C}, \hat{B}_1^+] = 1$, respectively.

As the final remark we notify that the normalizability of all the states obtained in our treatment depends crucially on the convergence of the series which define the normalization constants [Eqs. (4),(24-b),(41-b),(42-b),(51-b),(54-b)]. This means that the normalization constants should be nonzero and finite. This condition yields to the following restrictions for the values of $|z|^2 = (Kg\tau)^2 \rho_{aa}\rho_{bb}$ for each of the NLCSs,

$$|z\rangle_f \text{ [Eq.(2)]}: \quad |z|^2 < \lim_{m\to\infty}(m+1)f^2(m+1), \tag{55-a}$$

$$|z\rangle_f' \text{ [Eq.(23)]}: \quad |z|^2 < \lim_{m\to\infty}(m+1)/f^2(m+1), \tag{55-b}$$

$$|z,0\rangle_f \text{ [Eq.(41-a)]}: \quad |z|^2 < \lim_{m\to\infty}(m+1)/(4m+2)f^2(2m+2), \tag{55-c}$$

$$|z,1\rangle_f \text{ [Eq.(42-a)]}: \quad |z|^2 < \lim_{m\to\infty}(m+1)/(4m+6)f^2(2m+3), \tag{55-d}$$

$$|z,0\rangle_f^{(e)} \text{ [Eq.(51-a)]}: \quad |z|^2 < \lim_{m\to\infty}(2m+2)(2m+1)f^2(2m+2), \tag{55-e}$$

$$|z,1\rangle_f^{(o)} \text{ [Eq.(54-a)]}: \quad |z|^2 < \lim_{m\to\infty}(2m+3)(2m+1)f^2(2m+3). \tag{55-f}$$

Of course, depending on the form of $f(m)$, the range of $|z|^2$ may be unrestricted. For example if $f(m)$ decreases faster than $m^{-1}$ for large $m$, then in the cases of $|z\rangle_f'$, $|z,0\rangle_f$ and $|z,1\rangle_f$ the range of $|z|^2$ is unrestricted . While if $f(m)$ increases faster than $m$ for large $m$, then in the cases of $|z\rangle_f$, $|z,0\rangle_f^{(e)}$ and $|z,1\rangle_f^{(o)}$ there is no restriction on the range of $|z|^2$.

## 4. Summary

Our main work in this paper is investigating how to produce generic NLCSs (f-deformed CSs) in an ideal micromaser cavity which is pumped by a stream of two-level atoms prepared in a coherent superposition of the upper and lower states. Considering the intensity-dependent interaction, we have studied the quantum evolution of the cavity-field coupled to the polarized injected atoms through one as well as two-photon transitions. In the case of one-photon transition, two different types of NLCSs can be generated if the cavity-field starts from the vacuum state. On the other hand, we have found that in the case of two-photon transition four different types of NLCSs including nonlinear squeezed vacuum state, nonlinear squeezed first excited state, nonlinear even coherent state and nonlinear odd coherent state, can be generated. The presence of initial atomic coherence, weakness of atom-field coupling and largeness of passed atoms through the cavity are some of the essential assumptions in our treatment.